\documentclass[12pt,preprint]{emulateapj}

\usepackage{amsmath}
\usepackage{bm}

\slugcomment{\today}

\shorttitle{Mass and Radius of Neutron Stars and Supernova Neutrinos}
\shortauthors{Nakazato and Suzuki}

\begin{document}

\title{A New Approach to Mass and Radius of Neutron Stars with Supernova Neutrinos}

\author{Ken'ichiro Nakazato\altaffilmark{1}, and Hideyuki Suzuki\altaffilmark{2}}

\email{nakazato@artsci.kyushu-u.ac.jp}

\altaffiltext{1}{Faculty of Arts \& Science, Kyushu University, 744 Motooka, Nishi-ku, Fukuoka 819-0395, Japan}
\altaffiltext{2}{Faculty of Science \& Technology, Tokyo University of Science, 2641 Yamazaki, Noda, Chiba 278-8510, Japan}

\begin{abstract}
Neutron stars are formed in core-collapse supernova explosions, where a large number of neutrinos are emitted. In this paper, supernova neutrino light curves are computed for the cooling phase of protoneutron stars, which lasts a few minutes. In the numerical simulations, 90 models of the phenomenological equation of state with different incompressibilities, symmetry energies, and nucleon effective masses are employed for a comprehensive study. It is found that the cooling timescale is longer for a model with a larger neutron star mass and a smaller neutron star radius. Furthermore, a theoretical expression of the cooling timescale is presented as a function of the mass and radius and it is found to describe the numerical results faithfully. These findings suggest that diagnosing the mass and radius of a newly formed neutron star using its neutrino signal is possible.
\end{abstract}

\keywords{Core-collapse supernovae --- Neutrino astronomy --- Neutron star cores --- Neutron stars --- Nuclear astrophysics --- Supernova neutrinos}

\section{Introduction}\label{sec:intro}
The mass and radius of neutron stars provide clues to revealing the properties of high-density nuclear matter that are characterized by the equation of state (EOS) \citep[][]{latpra16,ozelfre16,oertel17,bali19}. Recently, constraints on the mass--radius relation have been obtained from observations of X-ray bursts from neutron star binaries with low-mass companions \citep[][]{steiner10,ozel10} and the gravitational wave from a binary neutron star merger, GW170817 \citep[][]{LIGO18}. As a result, EOSs that predict large ($\gtrsim$13.5~km) radii of neutron stars are disfavored. Furthermore, very recently, \citet{nicer21} and \citet{nicer24} have obtained new insights for the mass and radius of PSR J0030+0451 using {\it NICER}, which is a soft X-ray telescope launched in June 2017.

As is well known, neutron stars are formed in core-collapse supernova explosions. In this process, a large number of neutrinos are emitted, which were actually detected for SN1987A \citep[][]{hirata87,bionta87}. At that time, mass estimations for the newly born neutron star were attempted by using the neutrino event number, which should correspond to the binding energy of the neutron star emitted as neutrinos \citep[][]{sato87b,latyah89}. Although there were large uncertainties owing to low statistics for SN1987A, a statistically significant constraint will be provided by the next Galactic core-collapse supernova \citep[][]{suwa19}.

In this paper, we investigate how the supernova neutrino signal depends on not only the mass but also the radius of a neutron star. For this purpose, we perform comprehensive simulations of the neutrino emission from a protoneutron star (PNS), which is a nascent compact remnant of a core-collapse supernova. About half of the neutrinos are emitted within the dynamical phase of a supernova core, and the others originate from the cooling stage of a PNS \citep[][]{self13a,mir16,mueller19}. In the latter phase, a PNS is almost hydrostatic but it is hot and lepton-rich. Since the entropy and lepton number are reduced by neutrino diffusion, a PNS evolves into a cold neutron star, which takes about a few minutes.

The cooling process and neutrino emission of a PNS have been studied for a few decades \citep[][]{burr86,suzuki94,pons99,robe12b}. In particular, the effect of the EOS is one of the central issues in PNS cooling \citep[][]{sumi95,pons99,robe12,came17,self19}, and even the phase transitions that lead to black hole formation have been examined \citep[][]{keil95,pons01b}. The impacts of inhomogeneous matter at subnuclear densities have also been discussed and it was reported that the neutrino luminosity is insensitive to inhomogeneous matter and is mainly determined by the high-density EOS \citep[][]{self18}. Nevertheless, there have been no systematic studies on the EOS dependence except for our previous paper \citep[][]{self19}. In this study, we extend our investigations to further variations of the EOS. Moreover, we present a formulation of the cooling timescale for PNSs, which depends on their mass and radius, in order to compare it with our numerical results.

\section{Setup}\label{sec:setup}
The numerical methods employed in our simulations of PNS cooling are basically described in our previous paper \citep[][]{self19}. Quasi-static evolutions of the PNS are computed by solving the Tolman--Oppenheimer--Volkoff (TOV) and neutrino transport equations. The neutrino transfer, for which we adopt the multigroup flux-limited diffusion scheme \citep[][]{suzuki94}, is responsible for the time evolutions of the entropy and lepton-number profiles. We treat the three species of neutrinos, $\nu_e$, $\bar \nu_e$, and $\nu_x$ ($=\nu_\mu=\bar \nu_\mu=\nu_\tau=\bar \nu_\tau$), that interact with matter. For the neutrino interactions, while most of the rates are taken from \citet{bruenn85}, we include neutrino pair processes via nucleon bremsstrahlung \citep[][]{suzuki93} and plasmon decay \citep[][]{kohyama86}. Further sophistication of the medium modifications of neutrino opacities \citep[][]{fischer16,came17}, such as the nucleon dispersion-relation dependence of the interaction rates, is deferred to a future work. In this study, we do not consider effects of additional mass accretion \citep[][]{burrows88} and convection \citep[][]{robe12} for simplicity.

The initial conditions of our PNS cooling are adopted from the numerical results of hydrodynamical simulations as in \citet{self13a}. We compute the core collapse, bounce, and shock propagation for the progenitor models of $15M_\odot$ and $40M_\odot$ in \citet{woosley95}, utilizing the numerical code of general relativistic neutrino-radiation hydrodynamics in spherical symmetry \citep[][]{sumi05}. In these computations, we employ the Togashi EOS \citep[][]{togashi17}. Then, we obtain profiles of the entropy and electron fraction inside the shock wave as functions of the baryon mass coordinate, which are adopted as the initial conditions of our PNS cooling. Note that, although spherically symmetric models do not yield explosions in general, the region behind the stalled shock wave at some point in time can be regarded as a PNS model after shock revival. For the progenitor model of $15M_\odot$, snapshots when the shock wave is stalled at the baryon mass coordinates of $1.47M_\odot$ and $1.62M_\odot$ are adopted and referred to as models 147a and 162a, respectively. For the $40M_\odot$ progenitor, snapshots when the shock wave is at $1.62M_\odot$ and $1.78M_\odot$ are referred to as models 162b and 178b, respectively. While the PNS models 162a and 162b have the same baryon mass, the initial conditions are different.

For our PNS cooling, we employ a series of phenomenological EOSs. For uniform nuclear matter at zero temperature, we express the energy per baryon as
\begin{equation}
w(n_b,Y_p) = w_0 + \frac{K_0}{18n_0^2} (n_b-n_0)^2 + S(n_b)\,(1-2Y_p)^2,
\label{eq:symeos}
\end{equation}
with the baryon number density $n_b$ and proton fraction $Y_p$. Here, the saturation density and saturation energy are set to $n_0=0.16$~fm$^{-3}$ and $w_0=-16$~MeV, respectively. The stiffness of symmetric nuclear matter, for which $Y_p=0.5$, is characterized by the incompressibility $K_0$. The density-dependent symmetry energy is written as
\begin{equation}
S(n_b) = S_0 + \frac{L}{3n_0} (n_b-n_0) + \frac{1}{n_0^2} \left( S_{00} - S_0 -\frac{L}{3} \right)(n_b-n_0)^2,
\label{eq:hsymene}
\end{equation}
with the coefficients of the symmetry energy $S_0$ and its density derivative $L$ at the saturation density. Here, the symmetry energy at the density of $2n_0$ becomes $S_{00}$: $S(2n_0) = S_{00}$. To construct the finite-temperature EOS, we include the contribution of thermal effects evaluated using the ideal Fermi gas model, where the nucleon effective mass is a free parameter. Furthermore, our EOS shares the subsaturation-density region containing inhomogeneous nuclear matter with the Shen EOS \citep[][]{shen11}. Note that our EOS is described in detail in our previous paper \citep[][]{self19}.

So far, the properties of nuclear matter in the vicinity of the saturation density have been probed in several terrestrial experiments. In particular, for symmetric nuclear matter, experimental data on isoscalar giant monopole resonance suggest that $K_0$ lies around 230~MeV or 250~MeV \citep[][]{shlomo06,gc18}. On the other hand, for asymmetric nuclear matter, the values of $S_0$ and $L$ inferred from nuclear binding energies are correlated \citep[][]{kort10,latlim13}. They have been indicated to be 30~${\rm MeV} \le S_0 \le 32$~MeV and 40~${\rm MeV} \le L \le 60$~MeV if other experimental constraints are also considered \citep[][]{tews17}. To approach the symmetry energy at supranuclear densities, observations of neutron stars are advantageous. According to \citet{zhali19}, the symmetry energy at $2n_0$ is constrained to $46.9\pm10.1$~MeV using the data of GW170817. On the basis of these constraints, we calculate the structure of cold neutron stars using our EOS models and select the models that have a reasonable mass--radius relation. Incidentally, some models are inappropriate owing to the maximum mass of neutron stars being considerably smaller than the observations \citep[][]{nanogra18,croma19}.

\begin{figure}
\plotone{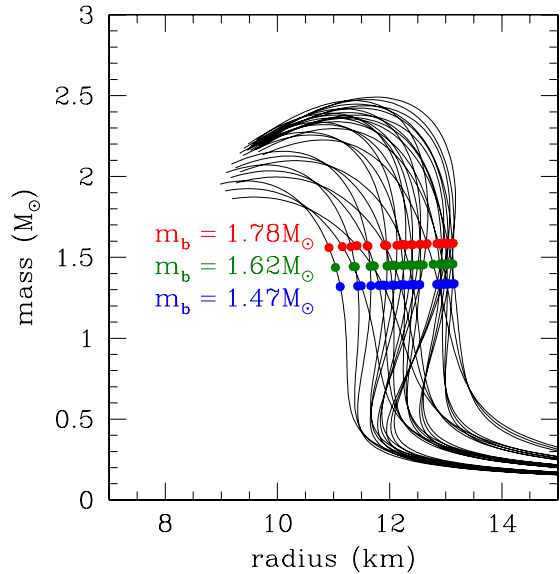}
\epsscale{1.0}
\caption{Mass--radius relations of cold neutron stars for our EOS models. Symbols denote the mass and radius of cold neutron stars with baryon masses of $1.47M_\odot$ (blue), $1.62M_\odot$ (green), and $1.78M_\odot$ (red).}
\label{fig:mr}
\end{figure}

In Figure~\ref{fig:mr}, we show the mass--radius relations of neutron stars for the EOS models selected in this study. Here, we examine the values of $K_0=220$, 245, and 270~MeV. For the symmetry energy $S(n_b)$, we employ the models with $(S_0,L,S_{00})=(30, 35, 35)$, $(30, 35, 40)$, $(30, 35, 45)$, $(30, 35, 55)$, $(31, 50, 40)$, $(31, 50, 45)$, $(31, 50, 55)$, $(32, 65, 45)$, $(32, 65, 55)$, and $(33, 80, 55)$, in units of MeV. Thus, for all possible combinations of $K_0$ and $(S_0,L,S_{00})$, we consider 30 models in total for the zero-temperature EOS. The neutron stars constructed in our EOS models have different radii ranging from 11~km to 13~km for typical masses. Nevertheless, the gravitational mass is insensitive to the EOS and is about $1.33M_\odot$, $1.44M_\odot$, and $1.57M_\odot$ for neutron stars with baryon masses of $1.47M_\odot$, $1.62M_\odot$, and $1.78M_\odot$, respectively (Figure~\ref{fig:mr}).

The effective mass of nucleons is a key ingredient in characterizing the finite-temperature EOS. It was reported that the likelihood of supernova explosion is increased for EOSs with a large effective mass \citep[][]{schne19}. In our EOS, the effective mass in units of nucleon rest mass is denoted by $u$ and we assume that neutrons and protons have the same value of $u$. Furthermore, we do not deal with the density and temperature dependences of the effective mass. Nevertheless, our EOS is advantageous for investigating the thermal contribution due to the effective mass because the dependences on the effective mass are separated from the variation of the zero-temperature EOS. In this study, we examine the cases with effective masses of $u=1$, 0.75, and 0.5 for each zero-temperature EOS. Thus, we employ 90 models of the finite-temperature EOS to perform the simulations of PNS cooling.

\section{Results and discussion}\label{sec:rad}
\begin{figure}
\plotone{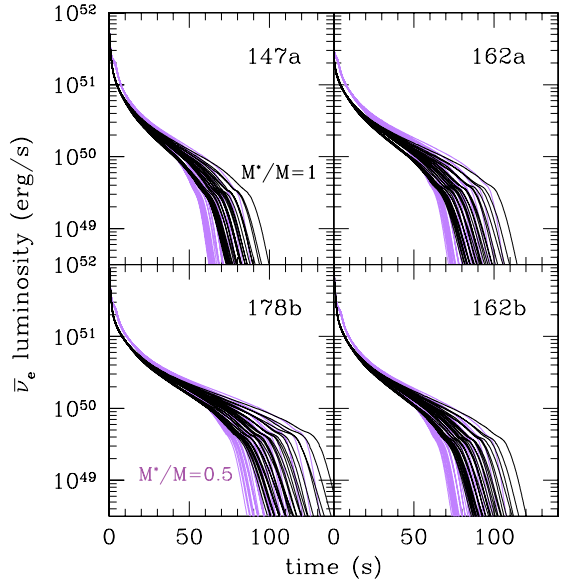}
\epsscale{1.0}
\caption{Luminosity of emitted $\bar\nu_e$ as a function of time for the PNS models of 147a (upper left), 162a (upper right), 162b (lower right), and 178b (lower left). The black and purple lines correspond to the cases with effective masses of $u=1$ and $u=0.5$, respectively.}
\label{fig:nulc}
\end{figure}

The light curves of $\bar\nu_e$, which are the time evolutions of neutrino luminosity, are shown in Figure~\ref{fig:nulc} for the models with effective masses of $u=1$ and $u=0.5$ The PNS models with a larger mass have a longer timescale of neutrino emission. This trend is qualitatively consistent with a previous study \citep[][]{came17}. The resultant neutrino light curves become similar for the models of 162a and 162b when the same EOS is adopted. This fact means that the neutrino signal in the late phase is insensitive to the initial profiles of the entropy and electron fraction \citep[][]{suwa19}. As reported in our previous paper \citep[][]{self19}, the decay timescale of the neutrino luminosity is longer for the models with larger effective masses because the thermal energy stored in the PNS is larger.

As shown in the neutrino light curves obtained from our computations (Figure~\ref{fig:nulc}), the PNS cooling is divided into three phases. Firstly, the neutrino luminosity decreases steeply due to the rapid contraction of the PNS because the neutrino luminosity is determined by the surface area. Secondly, since the structure of the PNS becomes almost stationary, the decrease of the neutrino luminosity becomes slower, which is identified as a shallow decay phase in the neutrino light curves. In this phase, the neutrinos trapped inside the PNS leak out from the surface gradually. The neutrino light curve of this phase is sensitive to the EOS \citep[][]{pons99,self19} and we focus on this phase hereafter. Finally, the matter in the PNS becomes neutrinoless $\beta$-equilibrium and the neutrino luminosity reduces steeply again.

\begin{figure*}
\plotone{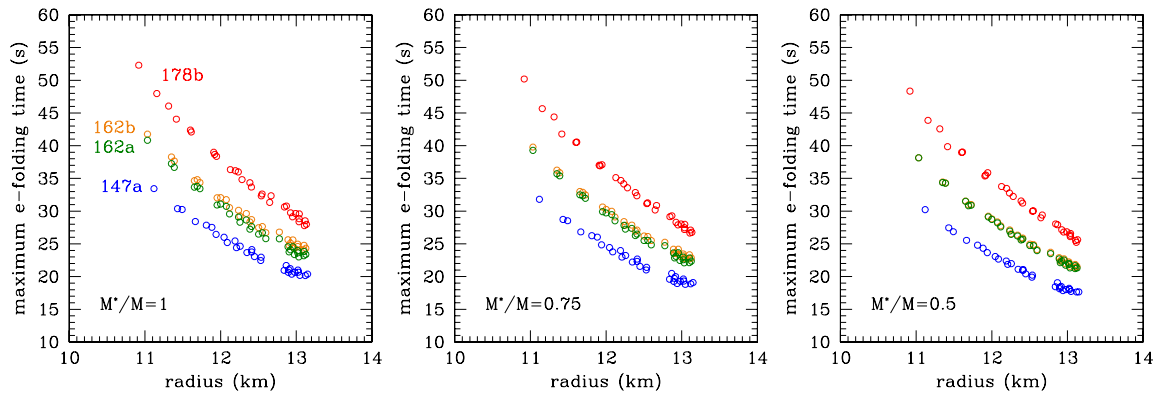}
\epsscale{1.0}
\caption{Maximum $e$-folding time of $\bar\nu_e$ luminosity as a function of the radius of the cold neutron star $r$ in the cases with an effective mass of $u=1$ (left), $u=0.75$ (center), and $u=0.5$ (right). Blue, green, orange, and red symbols correspond to the PNS models of 147a, 162a, 162b, and 178b, respectively.}
\label{fig:emax}
\end{figure*}

In this study, the $e$-folding time of the $\bar \nu_e$ luminosity, $\tau_{\bar\nu_e}(t)$, is introduced as
\begin{equation}
L_{\bar\nu_e}(t+\tau_{\bar\nu_e}(t)) = \frac{L_{\bar\nu_e}(t)}{e},
\label{eq:efold}
\end{equation}
where $L_{\bar\nu_e}$ is the $\bar \nu_e$ luminosity and $e$ is the base of the natural logarithm. Furthermore, we define the cooling timescale of a PNS as the maximum value of $\tau_{\bar\nu_e}$:
\begin{equation}
\tau_{\rm cool} \equiv \max_t \tau_{\bar\nu_e}(t).
\label{eq:tcool}
\end{equation}
In Figure~\ref{fig:emax}, $\tau_{\rm cool}$ is plotted as a function of the radius of the cold neutron star for each EOS model. Here, we can confirm that, whereas the PNS cooling is computed for various EOS models with different incompressibilities and symmetry energies, the cooling timescale is significantly correlated with the radius of the cold neutron star, and it is longer for an EOS model with a smaller neutron star radius.

\begin{figure}
\plotone{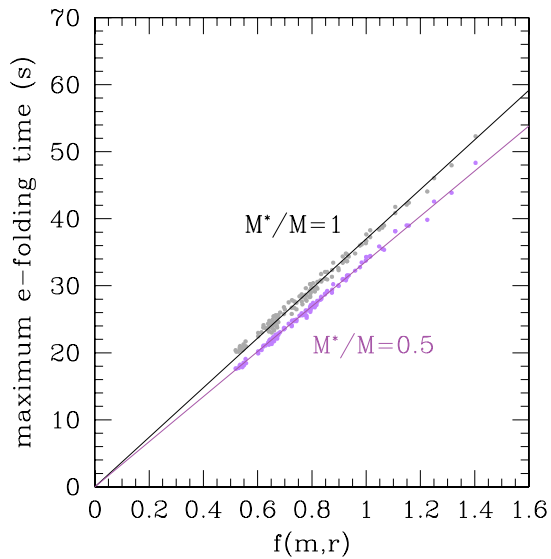}
\epsscale{1.0}
\caption{Maximum $e$-folding time of $\bar\nu_e$ luminosity as a function of $f(m,r) \equiv (m/1.4M_\odot)^2(r/10\,{\rm km})^{-3}(1-0.5\beta)^{-1}(1-2\beta)^{-1/2}$. Gray and purple symbols correspond to the cases with effective masses of $u=1$ and $u=0.5$, respectively. Lines show the theoretical formula of Eq. (\ref{eq:taufit}).}
\label{fig:nslp}
\end{figure}

As described above, the cooling timescale is longer for a model with a larger neutron star mass and a smaller neutron star radius. This feature can be formulated as bellow. The Kelvin-Helmholtz timescale is given by $\tau_{\rm KH} \approx |E_g|/L_\ast$, where $E_g$ is the stellar gravitational binding energy and $L_\ast$ is the stellar luminosity \citep[][]{kw90}. Since the luminosity is proportional to the stellar surface area, we can suppose
\begin{equation}
\tau_{\rm KH} \propto \frac{|E_g|}{R^2},
\label{eq:tkh}
\end{equation}
where $R$ is the stellar radius. In the Newtonian case, it is rewritten as $\tau_{\rm KH} \propto M^2/R^3$ because of $|E_g|\propto M^2/R$, where $M$ is the stellar mass. Thus the timescale is longer for stars with larger mass and smaller radius. However, for extending it to the neutron star with the mass of $m$ and radius of $r$,  we consider the following two relativistic effects. Firstly, a factor of $1/\sqrt{1-2\beta}$ is multiplied due to the time dilation being $\beta=Gm/rc^2$, where $c$ and $G$ are the velocity of light and the gravitational constant, respectively. Secondly, $|E_g|\propto M^2/R$ is replaced by the relativistic binding energy of the neutron star $E_b$. Therefore, assuming that the cooling timescale of PNSs is evaluated by the Kelvin-Helmholtz timescale, we obtain
\begin{equation}
\tau_{\rm cool} \propto \frac{E_b}{r^2\sqrt{1-2\beta}},
\label{eq:trkh}
\end{equation}
instead of Eq.~(\ref{eq:tkh}). According to \citet{lp01}, the binding energy of neutron stars is approximated for a large class of EOSs as
\begin{equation}
\frac{E_b}{m} = \frac{0.6\beta}{1-0.5\beta}.
\label{eq:lp01}
\end{equation}
Substituting Eq.~(\ref{eq:lp01}) into Eq.~(\ref{eq:trkh}), we can express the cooling timescale as
\begin{equation}
\tau_{\rm cool} = \tau^{\ast} \left( \frac{m}{1.4M_\odot} \right)^2 \left( \frac{r}{10\,{\rm km}} \right)^{-3}\frac{1}{(1-0.5\beta)\sqrt{1-2\beta}},
\label{eq:taufit}
\end{equation}
with a coefficient $\tau^{\ast}$.
Actually, as shown in Figure~\ref{fig:nslp}, we find that the cooling timescale of PNSs obtained from our simulations can be faithfully described by Eq.~(\ref{eq:taufit})
with $\tau^{\ast}=37.0$~s, 35.2~s, and 33.7~s in the cases with effective masses of $u=1$, 0.75, and 0.5, respectively.

\begin{figure}
\plotone{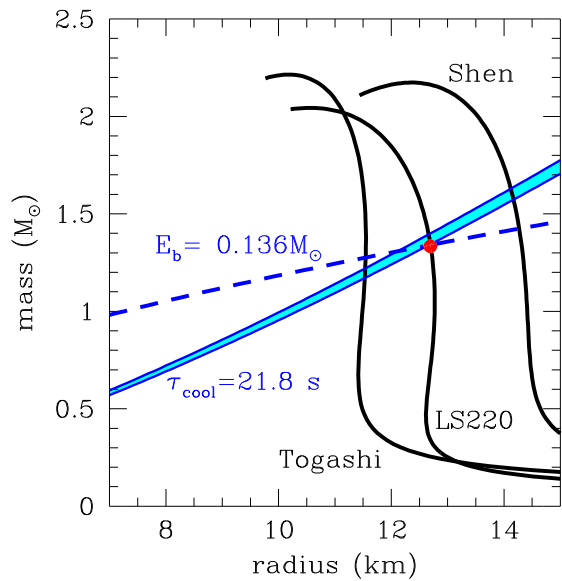}
\epsscale{1.0}
\caption{Neutron star mass and radius obtained from Eqs. (\ref{eq:lp01}) and (\ref{eq:taufit}). Dashed line and band respectively correspond to Eqs. (\ref{eq:lp01}) and (\ref{eq:taufit}) for the examples of the PNS models with a baryon mass of $1.47M_\odot$ and the LS220 EOS. Red symbol denotes $m$ and $r$ for the cold neutron stars corresponding to this example and black lines are the mass--radius relations of the Togashi, LS220 and Shen EOSs, from left to right.}
\label{fig:snns}
\end{figure}

If the cooling timescale is measured by the future detection of supernova neutrinos, it will be useful to probe the mass and radius of a newly formed neutron star. For example, in our previous paper \citep[][]{self19}, we have computed the PNS cooling of the model with a baryon mass of $1.47M_\odot$ for the EOS model proposed by \citet{ls91} with an incompressibility of 220~MeV (LS220 EOS), and we have obtained $\tau_{\rm cool}=21.8$~s. The relation between $m$ and $r$ given by Eq. (\ref{eq:taufit}) for this case is shown in Figure~\ref{fig:snns} with the mass--radius relation of neutron stars for the LS220 EOS. The intersection of them is consistent with the mass and radius of this model (shown by the red point). Furthermore, this model has $E_b=0.136M_\odot$ and the relation between $m$ and $r$ given by Eq. (\ref{eq:lp01}) is also shown in Figure~\ref{fig:snns}. Since $E_b$ corresponds to the total emission energy of supernova neutrinos, both of $\tau_{\rm cool}$ and $E_b$ are measurements of supernova neutrinos. Thus, by combining Eqs.~(\ref{eq:lp01}) and (\ref{eq:taufit}), we may be able to obtain constraints on the mass and radius of the neutron star from the neutrino observation.

In practice, some uncertainties should be considered to estimate the mass and radius of a neutron star applying our method. In this study, the effective mass of nucleons, which also affects neutrino opacities in the medium, is treated as an unknown parameter. As seen in the difference between the models 162a and 162b (Figure~\ref{fig:emax}), the cooling timescale has a slight dependence on the initial condition for the case with the large effective mass. Note that the initial condition of the model 162b has a higher entropy than that of the model 162a. While the entropy is originally generated by the shock propagation in the supernova core, the convection smooths the entropy gradient especially in the early phase \citep[][]{robe12}. Therefore, the convection would affect the cooling timescale as well as the initial entropy. While we define $\tau_{\rm cool}$ with the luminosity of $\bar \nu_e$, $E_b$ is the total emission energy of all neutrino flavors. Since energy equipartition among different flavors may not be achieved \citep[][]{mueller19}, different types of detectors are needed to evaluate the emission energy for each flavor \citep[][]{laha14,nik18,hlli18}. In any case, statistical uncertainties would be included when the supernova neutrinos are actually detected. Then, for reducing uncertainties, other constraints on the mass and radius of neutron stars, such as a mass--radius relation predicted by nuclear physics, are desired to be available. Therefore, various approaches are certainly worth investigation.

\section{Conclusion}\label{sec:conc}
In this paper, we have carried out a comprehensive simulation study of PNS cooling using 90 EOS models with different incompressibilities, symmetry energies, and nucleon effective masses. We have found that if the PNS mass is fixed, the cooling timescale depends on the radius of the cold neutron star in the final state and the nucleon effective mass, which is introduced to characterize the thermal properties of the EOS. Furthermore, we have presented a theoretical expression of the cooling timescale that describes our numerical results faithfully.

The findings in this study suggest that, as well as the total energy of emitted neutrinos, the decay timescale will be useful to probe the mass and radius of a newly formed neutron star. Actually, in our results for the PNS cooling timescale, the dependence on the zero-temperature EOS is encapsulated in a single parameter, the radius of the cold neutron star $r$, while we have examined various EOS models. In general, since the integration and slope of a neutrino light curve should have different dependences on $m$ and $r$, they can provide a complementary information as demonstrated in Figure~\ref{fig:snns}. For predicting $m$ and $r$ more accurately than this study, some sources of uncertainties, such as nucleon effective masses, neutrino opacities, and convection, need to be considered. However, this paper will provide an important and reliable basis for future work because Eq. (\ref{eq:taufit}) does not depend on details of the cooling model.

\acknowledgments
The authors are grateful to Ken'ichi Sugiura, Kohsuke Sumiyoshi, Yudai Suwa, and Shoichi Yamada for valuable comments. In this work, numerical computations were partially performed on the supercomputers at Research Center for Nuclear Physics (RCNP) in Osaka University. This work was partially supported by JSPS KAKENHI Grant Numbers JP26104006, JP17H05203, JP19H05802, and JP19H05811.

\end{document}